\documentclass[reqno]{amsart}
\usepackage{amsmath, amsthm ,amssymb,color}
\usepackage[mathscr]{euscript}
\usepackage{stmaryrd}
\usepackage{graphicx}
\usepackage{todonotes}
\usepackage[pdftex,breaklinks,colorlinks,
citecolor=blue,
urlcolor=blue,
pdftitle={Nematic first order phase transition for liquid crystals in the van der Waals--Kac
limit},
pdfauthor={C. Erignoux and A Giuliani}]{hyperref}
\usepackage{amsfonts, amscd, epsfig, amsmath, amssymb,enumerate}
\usepackage{graphicx}
\usepackage{color}
\usepackage{mathrsfs}

\usepackage{tikz}
\usetikzlibrary{backgrounds}
\usetikzlibrary{patterns,fadings}
\usetikzlibrary{arrows,decorations.pathmorphing}
\usetikzlibrary{decorations.pathreplacing}
\usetikzlibrary{decorations}
\usetikzlibrary{calc}
\usetikzlibrary{shapes.misc}
\definecolor{light-gray}{gray}{0.95}

\usepackage{float}
\def\centerarc[#1](#2)(#3:#4:#5){\draw[#1] ($(#2)+({#5*cos(#3)},{#5*sin(#3)})$) arc (#3:#4:#5);}

\allowdisplaybreaks 

\makeatletter
\@addtoreset{equation}{section}
\makeatother

\newtheorem{theorem}{Theorem}[section]

\newtheorem{remark}[theorem]{Remark}

\newcommand{\bb}[1]{{\mathbb #1}}
\newcommand{\bs}[1]{{\boldsymbol #1}}

\newcounter{as}[section]

\newcommand{\R}{{\mathbb R}}

\newcommand{\bfn}{{\bs n}}
\newcommand{\bfm}{{\bs m}}

\newcommand{\bfM}{{\bs M}}

\newcommand{\Z}{{\bb Z}}

\renewcommand{\S}{\mathbb{S}}

\newcommand{\N}{{\mathbb N}}

\newcommand{\pa}[1]{\left(#1 \right)}

\newcommand{\cro}[1]{\left[#1\right]}

\newcommand{\bfN}{{\bs{N}}}

\newcommand{\ccl}[1]{{\color{red}#1}}

\definecolor{dkgreen}{rgb}{0,0.6,0}
\definecolor{gray}{rgb}{0.5,0.5,0.5}
\definecolor{header}{gray}{0.3}

\author{Cl\'ement Erignoux}
\email{clement.erignoux@inria.fr}
\address{Equipe PARADYSE, Centre INRIA Lille Nord-Europe, Park Plaza, Parc scientifique de la Haute-Borne, 40 Avenue Halley B\^a timent B, 59650 Villeneuve-d’Ascq France}
\author{Alessandro Giuliani}
\email{giuliani@mat.uniroma3.it }
\address{Dip. Matematica e Fisica, Universit\`a Degli Studi Roma Tre, Largo San Leonardo Murialdo, 1 – 00146 Roma, Italy}
\address{Centro Linceo Interdisciplinare Beniamino Segre, Accademia Nazionale dei Lincei, Palazzo Corsini, Via della Lungara 10, 00165 Roma, Italy}

\title[Nematic 1st order phase trans. for liq. crystals in the VDW-K lim.]{Nematic first order phase transition for liquid crystals in the van der Waals--Kac
limit.}
\begin{document}
\maketitle

\begin{abstract}
In this paper we revisit and extend some mathematical aspects of Onsager's theory of liquid crystals that have been investigated in recent years by different communities
(statistical mechanics, analysis and probability). We introduce a model of anisotropic molecules with three-dimensional orientations interacting via a Kac-type interaction. We prove that, in the limit in which the range of the interaction is sent to infinity after the thermodynamic limit, the free energy tends to the infimum of an effective energy functional \`a la Onsager. We then prove that, if the spherical harmonic transform of the angular interaction has a negative minimum, this effective free energy functional displays a first order phase transition as the total density of the system increases.
\end{abstract}

\section{Introduction}
The problem of understanding phase transition phenomena in liquid crystals was first mathematically formalized in a seminal paper by Onsager \cite{Onsager}, in 
which he considered a simple microscopic model of anisotropic molecules interacting through repulsive interactions, and derived an effective free energy functional 
for the system, of the following form: 
\begin{equation}\label{OnsFun}
\frac{\rho}{\beta}\int_{\S^2} f(\Omega) \log  f(\Omega)   d\Omega +\frac{\rho^2}{2} \int_{\S^2\times \S^2} f(\Omega)f(\Omega')\varphi(\Omega\cdot\Omega')d\Omega d\Omega'. \end{equation}
Here $f(\Omega)$ represents the distribution of the orientation of the molecules, and $\varphi$ is an interaction potential, even  under 
`orientation flip', $\Omega\cdot\Omega' \to -\Omega\cdot\Omega'$. 
Since then, the topic of phase transitions in liquid crystal models has attracted significant attention, and the equilibrium phenomenology of liquid 
crystals is now well understood, see \cite{DP} for an extensive overview. 

However, from a more fundamental point of view, several aspects of the mathematical 
theory of liquid crystalline phases are not well understood. In particular, the problem of proving the existence of 
an oriented phase at low enough temperatures, or high enough densities, in a system of anisotropic molecules with continuous orientational 
symmetry and finite range interactions, is almost completely open (the only exceptions we are aware of, see \cite{AZ,ARZ,Za}, concern discrete, reflection positive, 
models with an internal $O(N)$ degree of freedom; or continuous models which can be reduced to such discrete, reflection positive, systems, via correlation inequalities, see \cite{GTZ}). Even less is rigorously known about the order of the phase transition from the disordered to the nematic phase,
which is supposed to be of {\it first} order in great generality, at least if $\Omega\in\S^n$, with $n\ge 2$: loosely speaking, as soon as the molecules acquire a common orientation, 
the density is expected to have a jump (at fixed activity), due to the fact that oriented elongated molecules can pack more efficiently than un-oriented ones. 
We believe that the lack of rigorous results on the existence and nature of the phase transition in microscopic models for liquid crystals 
is related to the limited mathematical understanding of continuous symmetry breaking phenomena in system with short range interactions: 
with a very few exceptions \cite{Ba,FS82,KK}, the only available results on the subject are based on reflection positivity and, therefore, are not robust under perturbations of the microscopic 
Hamiltonian. Typically, the `natural' microscopic models of liquid crystals are not reflection positive and, therefore, there are not many techniques available for attacking the problem. 
One possible route that, in our opinion, has not been explored enough in the context of liquid crystals, is to consider finite range models obtained as perturbation of mean field ones, 
in the spirit of \cite{LMP,Presuttibook}. The techniques developed by \cite{LMP} in their standard form are suitable for studying problems with discrete symmetry breaking only; still, 
there is hope that a generalization thereof can be used to attack the more challenging problem of continuous symmetry breaking and liquid crystalline order.

Of course, in order for such a program to be feasible, the scaling limit of the finite range model to the mean field one, as well as the properties of the effective energy functional for the limiting mean field model, 
must be understood in a complete and quantitative form. Motivated by this, in this paper we revisit and extend some aspects of the mathematical theory of liquid crystals that have been investigated in recent years by different communities (statistical mechanics, analysis, probability). 
In short, we introduce a model of anisotropic molecules with long but finite range interactions 
(we denote the range by $\gamma^{-1}$) and 
give a simple proof of the following facts: 
(1) as the range $\gamma^{-1}$ tends to infinity, the free energy of the model tends to the minimizer of a free energy functional {\emph{\`a la Onsager}}, i.e., of the form \eqref{OnsFun}; (2) 
this effective free energy functional displays a first order phase transition as the total density of the system is varied from small to large values. 
Our results are not the first of this kind available in the literature, and some aspects of our proofs overlap with 
known techniques developed by different mathematical communities: 
for instance, in the proof of (1), we use a criterium put forward by Lebowitz and Penrose \cite{LP} to show spatial homogeneity of the critical points, and 
in the proof of (2) we apply an argument already used in \cite{KR78, Vollmer} for analyzing the nature of the phase transition 
of the spatially homogeneous effective energy functional. Nevertheless, since the topic at hand involves various mathematical and physics communities and 
previous results are delocalized in papers aimed at researchers with different backgrounds, 
we deem worthwhile to have in the same short article both parts of the problem solved in a concise way.

\medskip

\subsection{Previous results on the derivation of an effective free energy functional.}
In the standard mean-field scaling limit, $N$ particles enclosed in a container interact among each other via a potential of strength $N^{-1}$ and range comparable 
with the container itself. Recently, models for nematic liquid crystals in this scaling limit have been considered \cite{BG}, and the free energy proved to 
converge to an appropriate effective energy functional, of the same form as the one originally considered by Onsager. 
A priori, the predictions on the nature of the phase transition based on the limiting mean-field functional are not reliable for finite range models 
(an exception is the case of finite range systems in large dimension, where in some cases the mean field approximation can be rigorously justified, 
see \cite{BC} for a proof covering models of interest for liquid crystals). In order to better understand the connection between the limiting functional and 
finite-range models, it is important to clarify whether there are other limit procedures  
leading to the Onsager functional. Onsager himself, in his original article \cite{Onsager}, derives his effective energy functional by truncating the 
\emph{virial expansion} for the free energy at second order, an approximation that, unfortunately, is justified only at low enough densities (well below
the critical density beyond which the system is expected to enter a nematic phase); see \cite{JKT} for a recent analysis 
of the virial expansion applicable to systems of anisotropic molecules in the canonical ensemble; see also \cite{PVZ} for a recent critical 
discussion of Onsager's approximation and its range of applicability. A third approach to derive the effective equation of state from microscopic models,
standing somehow in between the two previous approximation schemes, is based on the so-called `van der Waals limit', or `Kac limit', 
which has been rigorously proved to produce the expected effective energy functional for several models of {\it isotropic} particles, see
 \cite{GP1,LP}. In the van der Waals--Kac limit, particles interact on typical range of order $\gamma^{-1}$ (instead of the size $L$ of the macroscopic container),
 where $\gamma$ goes to $0$ \emph{after} having taken the thermodynamic limit $L \to \infty$ at constant density $\rho=N/L^d$ (here $L^d$ is 
 the $d$-dimensional volume of a cubic box of side $L$). One of the purposes of this note is to adapt the methods of  \cite{GP1,LP}
 to models of nematic liquid crystals, thus extending the proof of convergence to an effective energy functional {\emph{\`a la Onsager}} 
 beyond 
the mean field analysis of  \cite{BG}. Let us conclude this subsection by remarking that the problem of computing the free energy 
for models of anisotropic molecules with continuous symmetry and `really short-range' interactions (i.e., interactions of range comparable with the size of 
the molecules themselves), in regimes of intermediate densities (potentially including the critical density for nematic phase transition)
is completely open. For recent progress in the case of anisotropic molecules with {\it discrete} orientations, see \cite{DG,DGJ} and references 
therein; it would extremely interesting to extend these results to `clock-models' of anisotropic molecules with 
several, but finite, allowed orientations, but this remains to be done. 

\subsection{Previous results on the minimizers of the effective free energy functional.}
In his seminal paper, Onsager proved that the effective energy functional \eqref{OnsFun} displays a phase transition from isotropic to nematic 
liquid, at least for certain simple reasonable choices of $\varphi$, most notably $\varphi(x)=\sqrt{1-x^2}$, which is the potential arising from 
the truncation of the virial expansion, in the case of rod-like molecules. When we say that  `the effective energy functional displays a phase transition', 
we mean that, while the minimizer of \eqref{OnsFun} is 
isotropic for $\rho$ small enough, it is peaked around a given (arbitrary) direction $\Omega_0$ for $\rho$ large enough. Given this, 
two natural questions arise:
\begin{itemize}
\item Can we compute or characterize the critical points of the free energy functional?
\item Can we determine the order of the phase transition from isotropic to nematic liquid?
\end{itemize}
The problem of determining the order of phase transition has been studied for several effective free-energy functionals similar to the Onsager one, e.g., 
for those arising in the mean field solution of the classical XY and Heisenberg models, see \cite{KN}, 
and for the McKean-Vlasov functional, see \cite{CP}. In the latter case, the authors gave necessary and sufficient conditions for the existence 
of a first order phase transition, and exhibited examples of specific models in their class for which a first order phase transitions can be proved; however, 
their analysis does not apply to the Onsager case. 

In order to attack both questions in the Onsager case, a natural approach is to study the Euler-Lagrange equation 
\begin{equation} \label{ELeq} f(\Omega)=\frac{ e^{- \beta\rho\int \varphi(\Omega\cdot \Omega')f(\Omega') d\Omega'}}{\int e^{ - \beta\rho\int \varphi(\Omega\cdot \Omega')f(\Omega')  d\Omega'}d\Omega}.\end{equation}
Unfortunately, for general interactions $\varphi$, including the case $\varphi(x)=\sqrt{1-x^2}$, this equation is infinite-dimensional and, therefore,
very hard to solve or analyze. Remarkably, there are special cases in which this equation reduces to a finite-dimensional one, most notably 
the case of the so-called Maier-Saupe potential \cite{Freiser}, $\varphi(x)=1-x^2$. For this potential,  in the case of three-dimensional orientations ($\Omega\in \S^2$),
the Euler-Lagrange equation was solved independently in \cite{FS} and \cite{LZZ}. These papers derive a complete classification of the critical points and 
bifurcation diagram. Although the issue of the order of the phase transition is addressed neither in  \cite{FS} nor in \cite{LZZ}, one can prove 
that the phase transition is discontinuous (first order); this follows, in particular, from our Theorem \ref{thm:PhaseTransition} below, see Section \ref{sec:applications}. 
For the `Onsager case' $\varphi(x)=\sqrt{1-x^2}$, 
Kayzer and Ravech\' e \cite{KR78} built an iterative scheme to compute the axially symmetric solutions of the Euler-Lagrange equation; recently, 
Vollmer \cite{Vollmer} obtained the full classification of the bifurcation points from the uniform solution. An extension of Vollmer's results, 
included in the present paper, implies that the phase transition is first order in the Onsager case, as well. 

Let us remark that an analogous bifurcation analysis for the Euler-Lagrange equation \eqref{ELeq} in the case of two-dimensional orientations ($\Omega\in \S^1$)
has been worked out in \cite{CLW10,NY18} for various potentials, including Maier-Saupe and Onsager's. 
The classification and characterization of the critical points has been obtained in \cite{FS2} for the family of potentials $\cos\big(n\,\theta(\Omega,\Omega')\big)$,
where $\theta=\theta(\Omega,\Omega')$ is the angle formed by $\Omega$ and $\Omega'$, and $n\geq 1$; for all these cases the phase transition is continuous 
(second order). Note that this family includes Maier Saupe potential ($n=2$); the general case
remains open, but we expect that the transition is generically continuous, see Remark \ref{rem:d2} below. 

\subsection{Plan of the paper}
In Section \ref{sec:main} we define the liquid crystal model we consider, and state our two main results, namely a variational formula for the thermodynamic free energy 
(Theorem \ref{thm:FreeEnergy}), and a simple criterion for the existence of a first order phase transition for the free energy (Theorem \ref{thm:PhaseTransition}). 
In Section  \ref{sec:applications}, we exhibit some concrete models, which these results apply to. In Section \ref{sec:GP}, we prove the variational formula for the free 
energy, by adapting the proofs by Lebowitz and Penrose \cite{LP} and Gates and Penrose \cite{GP1} on the van der Waals--Kac limit to our case of interest, where 
particles have an internal orientational degree of freedom. 
Finally, in Section \ref{sec:PhaseTrans}, we prove our criterium for the first order nature of the phase transition, 
and discuss the fundamental differences between models with two- and three-dimensional orientations. 

\section{Microscopic model and main results}
\label{sec:main}
We consider a system of infinitely thin rods interacting via a pair wise potential and hard-core repulsion, modelled as follows. 
Given $L>0$, $d\geq 1$, consider a large box $\Lambda_L=[0,L]^d$, containing $N$ anisotropic particles, each characterized by a position 
$x_i\in \Lambda_L$, to be thought of as its center, and a three-dimensional orientation $\Omega_i\in\S^2$. Note that we do not require the space dimension $d$ to be the same as the rod orientation's. Letting $\bar x=(x_1,\dots,x_N)$ and $\bar \Omega=(\Omega_1,\dots,\Omega_N)$, we assume the 
particles to interact via the pairwise potential
\[V_\gamma(\bar x,\bar \Omega)=\sum_{1\leq i<j\leq N}v_\gamma(x_i-x_j, \Omega_i, \Omega_j).\]
Denoting $\Omega\cdot\Omega'$ the inner product in $\R^3$, we assume the potential $v_\gamma$ to have the form 
\[v_\gamma(r, \Omega, \Omega')=q(r)+\gamma^d \varphi(\gamma r, \Omega\cdot\Omega'),\]
where $q(r)$ models an isotropic hard core repulsion with distance $r_0$, 
\[q(r)=\begin{cases}
\infty&\mbox{ if }|r|\leq r_0\\
0&\mbox{ otherwise}
\end{cases},\]
and $\varphi:\R^d\times [-1,1]\to\R$ models the anisotropic long range interaction (the inverse range $\gamma$ of $\varphi$ 
should be thought of as a small parameter). 
We assume that $\varphi$ is integrable on $\R^d\times [-1,1]$, and, more specifically, that 
\begin{equation} \sup_{\tau>0}\sup_{x\in\Lambda_\tau}\sup_{u\in[-1,1]} \tau^d\sum_{n\in\mathbb Z^d}\big|\varphi(x+\tau n,u)\big|<+\infty.
\label{eq:supsumbound}\end{equation}
Moreover, we assume that 
\begin{align}
&\mbox{$\varphi(x,\cdot)$ is $C_1(x)$-Lipschitz.}\label{eq:assphi}\\
&\mbox{$\varphi(\cdot,v)$ is continuously differentiable.} 
\label{eq:assphi2}\\
&\mbox{Both $C_1(x)$ and $C_2(x):=\sup_v|\nabla_x\varphi(x,v)|$ are Riemann integrable over $\R^d$.} \label{eq:assphi3}
\end{align}
The fact that the potential models a liquid crystalline interaction translates into $\varphi$ being even in its second variable. However, since this is not necessary to derive the free energy functional, we keep it as an assumption for our second result, which states that a first order phase transition occurs in the thermodynamic limit.

\medskip

The Gibbs distribution for this system with open boundary conditions is then given by
\[\mu_{\beta,\gamma}(\{X_i\in x_i+dx_i,\; O_i\in \Omega_i+d\Omega_i\}_{i=1,\ldots,N})=\frac{1}{Z_{ \beta}(N,L,\gamma)}\frac{d\bar x d \bar\Omega}{N!}e^{-\beta V_\gamma(\bar x , \bar \Omega)},\]
where  $Z_{ \beta}(N,L,\gamma)$ is the partition function 
\[Z_{\beta}(N,L,\gamma ):=\int_{\Lambda_L^N\times (\mathbb{S}^2)^N}\frac{d\bar x d\bar\Omega}{N!}e^{-\beta V_\gamma(\bar x, \bar \Omega)}.\]
The thermodynamic free energy in the van der Waals--Kac limit is defined as 
\[\mathcal{F}_\beta(\rho)=\lim_{\gamma\to 0}\lim_{L\to\infty}\frac{-1}{\beta L^d}\log Z_{\beta}(\lfloor\rho L^d\rfloor ,L,\gamma).\]

Our goals are: first, to derive an expression for the free energy as a variational principle over the distribution in space and orientation of the particles; then, to show that this expression undergoes a first order phase transition under suitable assumptions on the interaction potential $\varphi$.
To state our first result, denote by $\rho_{cp}=\rho_{cp}(r_0)$ the close packing density at radius $r_0$. Introduce the partition function relative to $N$ particles with hard core repulsion, 
\begin{equation}
\label{eq:defZ0}
Z_{hc}(N,L)=\int_{\Lambda_L^N}\frac{d\bar x }{N!}{\bf 1}_{\{x_i-x_j\geq r_0,\; \forall i\neq j\leq N\}},
\end{equation}
and define the corresponding free energy at density $\rho$ as 
\begin{align}
\mathcal{F}_{hc}(\rho)&=\lim_{L\to\infty}\frac{-1}{\beta L^d}\log Z_{hc}(\lfloor\rho L^d\rfloor ,L)\label{Fhc}\\
&=\begin{cases}
\beta^{-1}\big[\rho\log(\rho/e) +\mathcal{Q}_{r_0}(\rho)\big]&\mbox{ for }\rho<\rho_{cp}\\ 
\infty &\mbox{ otherwise}
\end{cases},\nonumber
\end{align}
where the {\it positive} term $\mathcal{Q}_{r_0}(\rho)$ encompasses the loss of entropy due to the hard core repulsion, \ccl{for which unfortunately no explicit formula can be derived}. It is well known, \ccl{however, that the limit \eqref{Fhc} is well defined (see e.g. \cite{COMPLETE}), and that} if 
$\rho r_0^d$ is sufficiently small, $\mathcal{Q}_{r_0}(\rho)$
is real analytic; moreover, both $\mathcal{Q}_{r_0}(\rho)$ and $\rho^{-1}\mathcal{Q}_{r_0}(\rho)$ are increasing in $r_0$, and convex and increasing in $\rho$ 
(these properties follow from the fact that the second and third virial coefficients of the hard core gas are positive, see \cite{LB82}).
We are now ready to state our main results.

\begin{theorem}
\label{thm:FreeEnergy}
Given a function $f:\R^d\times \S^2\to\R$, we denote by $\bar f$ the function on $\R^d$ defined as $\bar f(x)=\int_{\S^2}f(x,\Omega)d\Omega$. Then
\begin{equation} \mathcal{F}_\beta(\rho)=\inf_{\substack{ \tau\ge 0\\ 
f \in \mathscr{R}_\tau}}\mathcal{F}_{\beta,\tau}(\rho,f),\end{equation}
where 
\begin{eqnarray}
\mathcal{F}_{\beta,\tau}(\rho,f)&:=&\frac1{\tau^d}\bigg\{\frac{1}{\beta}\int_{\Lambda_\tau} \mathcal{Q}_{r_0} (\rho\bar f(x)) dx +\frac{\rho}{\beta}\int_{\Lambda_\tau\times \S^2} f \log  f(x,\Omega)  dx\, d\Omega \label{eq:freeenergyfun}\\
&+&\frac{\rho^2}{2} \int_{\Lambda_\tau\times \R^d\times \S^2\times \S^2} f(x,\Omega)f(y, \Omega')\varphi(x-y,\Omega\cdot\Omega')dx\, dy\, d\Omega\, d\Omega' \bigg\},\nonumber
\end{eqnarray}
and $\mathscr{R}_\tau$ is the set of non-negative, $L^1_{loc}$, $\tau$-periodic functions,
such that $$\tau^{-d}\int_{\Lambda_\tau\times\S^2} f(x,\Omega)dx\,d\Omega=1.$$
[If $\tau=0$, $\mathscr{R}_0$ consists of functions that are translationally invariant in $x$; if $f\in \mathscr{R}_0$, we simply denote by $f(\Omega)$
the values of $f$.]
\end{theorem}
\ccl{Note that the first two terms in\eqref{eq:freeenergyfun} represent the contribution of the hard core free energy $\mathcal{F}_{hc}$ defined in \eqref{Fhc} in the angle-dependent setting.} Since $\mathcal{Q}_{r_0}(\rho)=+\infty$ for $\rho\ge \rho_{cp}$, in the minimization of $\mathcal{F}_{\beta,\tau}(\rho,f)$ 
we can assume without loss of generality that $\rho\bar f(x)<\rho_{cp}$.  
For later reference, we denote $\mathcal{F}_\beta(\rho,f):=\mathcal{F}_{\beta,0}(\rho,f)$, and note that, for $f\in \mathscr{R}_0$, 
\begin{equation}\mathcal{F}_\beta(\rho,f)=\frac{1}{\beta} \mathcal{Q}_{r_0} (\rho)+\frac{\rho}{\beta}\int_{\S^2} 
f \log f(\Omega)   d\Omega +\frac{\rho^2}{2} \int_{ \S^2\times \S^2} f(\Omega) f(\Omega')\widehat\varphi(0,\Omega\cdot\Omega')d\Omega\, d\Omega', 
\nonumber\end{equation} 
where $\widehat\varphi(0,u)=\int_{\R^d} \varphi(x,u) dx$. 

\medskip

Let us now give a criterion for $\mathcal{F}_\beta(\rho)$ to exhibit a first order phase transition.
Since we are mostly interested in the angular dependence of the free energy functional, we focus on the $r_0\to 0^+$ limit of 
$\mathcal{F}_\beta(\rho)$, which we denote by $\mathcal{F}_\beta^0(\rho)$ (we define $\mathcal{F}_{\beta,\tau}^0(\rho,f)$
and $\mathcal{F}_{\beta}^0(\rho,f)$ analogously). Most of the consideration below can be extended to the case of $r_0$ small, thanks to the properties 
of $\mathcal Q_{r_0}$ spelled after \eqref{Fhc}, but we will not discuss this issue explicitly below. 

Denote by $P_\ell$, with $\ell\in\N_0$ (here $\N_0$ is the set of non-negative integers), 
the $\ell$-th Legendre polynomial, whose definitions and basic properties are briefly recalled in Appendix \ref{app:Legendre}.
\begin{theorem}
\label{thm:PhaseTransition}
For $\xi\in \R^d$, $\ell\in \N_0$ define 
\begin{equation}
\label{eq:defPhi}
\widehat{\Phi}_\ell(\xi)=\int_{\R^d\times[-1,1]} \varphi(x,u)P_\ell(u) e^{-ix\cdot\xi}du dx.\end{equation}
Assume that for any $x\in \R^d$, $\varphi(x,\cdot)$ is even, and that there exists $\ell^\star>0$ such that 
\begin{equation}
\label{eq:cond_phi}
\inf_{(\xi,\ell)\in \R^d\times \N }\widehat{\Phi}_\ell(\xi)=\widehat{\Phi}_{\ell^\star}(0)<0.
\end{equation}
Then $\mathcal{F}^0_{\beta}(\rho)$ 
exhibits a first-order phase transition, in the sense that there exists a positive critical density $\rho_c$, strictly smaller than 
$\rho^\star:=-4\pi/\big(\beta \hat \Phi_{\ell^\star}(0)\big)$, such that:
\begin{itemize}
\item [1.] For any $\rho<\rho_c$, the uniform profile $f_0\equiv \frac{1}{4\pi}$ is the unique global minimizer of $\mathcal{F}^0_{\beta,\tau}(\rho, f)$, for 
all $\tau\ge 0$. 
\item [2.] For any $\rho_c<\rho<\rho^\star$, the uniform profile $f_0$ is a local stable minimum of $\mathcal{F}^0_{\beta,\tau}(\rho, f)$, for 
all $\tau\ge 0$; however, there exists $\tau\ge 0$ and $f'\in \mathscr{R}_\tau$ such that 
$\mathcal{F}^0_{\beta,\tau}(\rho, f')<\mathcal{F}_\beta^0(\rho, f_0)$.
\item [3.] For any $\rho>\rho^\star$, the uniform profile is locally unstable.
\end{itemize}
\end{theorem}
\begin{remark}[On the parity of $\ell^\star$]
\label{rem:elleven}
Since the Legendre Polynomials $P_\ell$ are even (resp. odd) on $[-1,1]$ iff $\ell$ is, and since we assumed our potential $\varphi$ to be even in its second variable,
we obtain by symmetry $\widehat{\Phi}_\ell\equiv 0$ for any odd $\ell$. In particular, one must have, under the assumptions of 
Theorem \ref{thm:PhaseTransition}, that $\ell^\star$ is even.
\end{remark}

The proof of Theorem \ref{thm:FreeEnergy} mimicks the one of \cite{GP1} and is given in Section \ref{sec:GP}. The proof of Theorem \ref{thm:PhaseTransition} 
is given in Section \ref{sec:PhaseTrans} and goes as follows. 
In our setting of pairwise particle interactions, the energetic contribution to the free energy can be expressed as a quadratic functional of the particles distribution, both in space and orientation. In this context, one can then develop the entropic term around the uniform profile $f_0$ to second order to determine the density $\rho^\star$ at which $f_0$ loses linear stability, as a function of the most negative eigenvalue of the energetic contribution. Developing further, two cases can arise: either at $\rho^\star$ the leading correction beyond the quadratic approximation is negative, in which case at the critical density $\rho^\star$ the uniform profile is not a local minimizer for the free energy, which is sufficient to prove a first order phase transition; or the leading order term is positive, in which case the uniform profile is a local minimizer. For general liquid crystal models with orientation in $\S^2$, the first case arises and a first order phase transition can then be proved. 
Notably, this is not the case for ferromagnetic models (cf. Remark \ref{rem:ferromag} below) or two-dimensional liquid crystals for which the orientation is in $\S^1$ (cf. Remark \ref{rem:d2} below). In both of those cases, at the density $\rho^\star$, the uniform profile $f_0$ is, at least locally, a minimizer of the free energy. This, of course, is not sufficient to preclude the existence of a first order phase  transition, but still sheds some light on the different phenomenologies of these three closely related models.

\section{Examples of applications}
\label{sec:applications}
Before proving Theorem \ref{thm:FreeEnergy} and Theorem \ref{thm:PhaseTransition}, let us exhibit some explicit models which they can be applied to. 
Assume for simplicity  that $\varphi$ has separate variables, 
\[\varphi(x,u)=\phi(x)g(u),\]
with $\phi:\R^d\to \R$, $g:[-1,1]\to\R$. In this case, defining $\lambda_\ell:=\int P_\ell(u)g(u)du$, the assumptions required for Theorems \ref{thm:FreeEnergy} and \ref{thm:PhaseTransition} to hold translate into 
\begin{eqnarray}
&&\phi\mbox{ is positive definite, $C^1$, and integrable over $\R^d$, together with its derivative},\nonumber\\
&&g\mbox{ is even, Lipschitz and $\exists \ell^\star>0$, s.t. $\inf_{\ell>0} \lambda_\ell =\lambda_{\ell^{\star}}<0$}.\nonumber
\end{eqnarray}
Without loss of generality, we also assume that $\int_{\R^d}\phi(x)dx=1$. 

The case of Onsager's potential corresponds to the choice $g(u)=\sqrt{1-u^2}$. In this case, the $\lambda_{\ell}$'s can be explicitly computed and form for $\ell\geq 1$ an increasing sequence. Indeed, denoting by $P_\ell^m$ the associated Legendre polynomials,
see \eqref{assLegpol}, 
\begin{equation*}
\lambda_{\ell}=\int_{-1}^1\sqrt{1-v^2}P_\ell(v)dv
=\frac{1}{2\ell+1}\int_{-1}^1\big(P^1_{\ell-1}(v)-P_{\ell+1}^{1}(v)\big)dv .
\end{equation*}
For $m=1$, (cf. \cite{INTALP}, p.646)
\[R^1_{2\ell+1}:=\int_{-1}^1P_{2\ell+1}^1(u)du=-\frac{\pi(2\ell+2)}{(2\ell+1)2^{4\ell+4}}\binom{2\ell+2}{\ell+1}^2\]
so that 
\begin{align*}\lambda_{2\ell}&=\frac{1}{4\ell+1}\pa{R^1_{2\ell-1}-R^1_{2\ell+1}}=-\frac{\pi}{2(\ell+1)(2\ell-1)2^{4\ell}}\binom{2\ell}{\ell}^2.\end{align*}
Note that this sequence is non-decreasing for positive indexes and vanishes as $\ell\to\infty$. In particular, $\lambda_\ell^\star=\lambda_2=-\pi/16$. 
Therefore, by Theorem \ref{thm:PhaseTransition}, the system undergoes a first order phase transition,  which occurs before the uniform profile $f_0$ loses linear stability at $\rho^\star=64/\beta$. 

\medskip

The same explicit characterization of the phase transition for the limiting functional can be extended to more general functions $g$ satisfying the assumptions above, provided we can compute the smallest $\lambda_\ell:=\int P_\ell(u)g(u)du$ and show that it is negative. For example, choosing  $g=(1-u^2)^k$, corresponding to an interaction $\big[\sin \big(\theta(\Omega, \Omega')\big)\big]^{2k}$, the corresponding $\lambda_\ell^{(k)}$ can be computed explicitly, recursively in $k$, thanks to the identity
\[(1-u^2)P_\ell=\frac{(\ell+1)(\ell+2)}{(2\ell+1)(2\ell+3)}(P_\ell-P_{\ell+2})-\frac{\ell(\ell-1)}{(2\ell+1)(2\ell-1)}(P_{\ell-2}-P_\ell),\]
which follows from using three times Bonnet's recursion formula \eqref{eq:Bonnet}. By this identity, the computation of $\lambda_\ell^{(k)}$ 
can be reduced to that of $\lambda_{\ell}^{(k-1)}$, $\lambda_{\ell-2}^{(k-1)}$ and $\lambda_{\ell+2}^{(k-1)}$. For Maier-Saupe's potential, corresponding to $k=1$, one obtains that $\lambda_0=4/3$,  $\lambda_{2}=-4/15$, and $\lambda_\ell=0$ for any other $\ell>0$, so that the loss of linear stability occurs at $\rho^\star=15\pi/\beta$.
Solving the recursion equation above, one can analogously derive the threshold for the linear stability of the homogeneous profile for larger values of $k$.

\section{Thermodynamic limit: Proof of Theorem \ref{thm:FreeEnergy}.}
\label{sec:GP}

As anticipated above, to prove Theorem \ref{thm:FreeEnergy}, we follow \cite{GP1}. 
For the sake of conciseness, we will not detail some technical steps already solved in \cite{GP1}, and instead present the general structure of the proof and focus on 
the necessary modifications to account for the presence of the angular variables $\Omega_i$.  
Throughout, we tesselate $\Lambda_L$ into $\bfn:=n^d$ boxes $\Lambda_\ell^i,$ of side $\ell$, for   $1\leq i\leq \bfn$ where $n=L/\ell$. We further consider the inner box $\hat\Lambda_\ell^i \subset \Lambda_\ell^i$, with same center as $\Lambda_\ell^i$ and side $k=k_\ell:=\ell-\sqrt{\ell}$ instead of $\ell$. We also tesselate $\S^2=\{\Omega=(\theta, \phi)\in [0,\pi]\times[0, 2\pi)\}$ into $\bfm:= m^2$ pieces  $\S_i$ each of surface $s:=4\pi/\bfm$. To give the reader a sense of the relative scales of those parameters, to carry out the proof, we will consider the system in the limit
\begin{equation} \label{eq:quadruplelimit}L\to\infty, \mbox{ \emph{then} }\gamma \to 0, \mbox{ \emph{then} }\ell\to \infty\mbox{ \emph{then} }m\to\infty.\end{equation}

\subsection{Upper bound}
We first investigate the upper bound for the free energy, which corresponds to a lower bound on the partition function. 
Fix a family of integers $\bfN=(N_i^p)_{1\leq i \leq \bfn, \; 1\leq p\leq \bfm}$ satisfying $\sum_{i,p} N_i^p=N$, and define $N_i=\sum_{p}N_i^p$. We can then write 
\[Z_{\beta}(N,L,\gamma )\ge \sup_{\bfN}\left\{\frac{1}{\prod_{i,p}N_i^p!}\int_{(\hat \Lambda^1_L\times \S_1)^{N_1^1}}\dots\int_{(\hat \Lambda^\bfn_L\times \S_\bfm)^{N_{\bfn}^\bfm}}d\bar x d\bar\Omega e^{-\beta V_\gamma(\bar x,\bar\Omega)}\right\}.\]
Because of the hard core repulsion, if $N_i>\ell^d \rho_{cp}$, the integrand vanishes, so that we can safely assume that each of the $N_i$'s is bounded by $\ell^d\rho_{cp}$.  Assume that $\ell$ is large enough so that  $\sqrt{\ell}>r_0$, so that particles in different boxes do not interact via the hard core interaction. Recalling from \eqref{eq:defZ0} the definition of the hard-core partition function $Z_{hc}(N_i,k)$ , we obtain
\begin{equation}
\label{eq:Zinf1}
Z_{\beta}(N,L,\gamma )\ge \sup_{\bfN}\left\{|\S_1|^{N}\frac{\prod_{i}N_i!}{\prod_{i,p}N_i^p!}\prod_{i}Z_{hc}(N_i,k) e^{-\beta W_{\max}(\bfN)}\right\},
\end{equation}
where $W_{\max}(\bfN)$ is the maximum of $V_\gamma(\bar x,\bar\Omega)$ over all $\bfN$ summing to $N$, such that $N_i\leq\rho_{cp} \ell^d$, and with $N_i^p$ particles in $\hat\Lambda_\ell^i\times \S_p$. Denoting
\begin{equation*}
\underline{\varphi}_{i,j}^{p,q}=\inf_{\substack{(x,x')\in \Lambda_\ell^i\times \Lambda_\ell^j\\(\Omega, \Omega')\in \S_p\times \S_q}}\varphi(\gamma(x-x'), \Omega\cdot\Omega')
\end{equation*}

\begin{equation*}
\overline{\varphi}_{i,j}^{p,q}=\sup_{\substack{(x,x')\in \Lambda_\ell^i\times \Lambda_\ell^j\\(\Omega, \Omega')\in \S_p\times \S_q}}\varphi(\gamma(x-x'), \Omega\cdot\Omega'),
\end{equation*} 
and noting that, for any $i$, $\varphi(\gamma(x_i-x_i),\Omega_i\cdot \Omega_i)=\varphi(0,1)$, 
we find
\begin{equation}
\label{eq:Z1}
W_{\max}(\bfN) =\frac{\gamma^d}{2} \sum_{\substack{i,j\leq \bfn\\
p,q\leq \bfm}}N_i^pN_j^q\underline{\varphi}_{i,j}^{p,q}-\frac{N \gamma^d}{2}\varphi(0,1)+\Delta(\bfN),
\end{equation}
where the supremum is taken over all families $N_i^p$ summing to $N$, and $\Delta(\bfN)$ is bounded by
\[|\Delta(\bfN)|\leq \frac{\gamma^d}{2} \sum_{\substack{i,j\leq \bfn\\
p,q\leq \bfm}}N_i^pN_j^q[\overline{\varphi}_{i,j}^{p,q}-\underline{\varphi}_{i,j}^{p,q}].\]
For any $\Omega\in \S^2$, and any $\Omega_1, \Omega_2\in \S_p$ for $p\leq \bfm$, we have $|\Omega\cdot(\Omega_1-\Omega_2)|\leq c_0/m$, 
for some constant $c_0>0$. Therefore, by triangular inequality we obtain 
\begin{multline*}
\overline{\varphi}_{i,j}^{p,q}-\underline{\varphi}_{i,j}^{p,q}\leq\sup_{u\in [-1,1]}\cro{\sup_{(x, x')\in \Lambda_\ell^i\times \Lambda_\ell^j} \varphi(\gamma(x-x'),u)-\inf_{(x, x')\in \Lambda_\ell^i\times \Lambda_\ell^j}\varphi(\gamma(x-x'),u)}\\
+\frac{2 c_0}{m}\sup_{(x, x')\in \Lambda_\ell^i\times \Lambda_\ell^j}C_1(\gamma (x-x')):=A_{i,j}+B_{i,j},
\end{multline*}
where $C_1(x)$ is the Lipschitz constant appearing in assumption \eqref{eq:assphi}.
The right hand side above no longer depends on $p,q$. In particular,
\begin{equation}|\Delta(\bfN)|\leq \frac{\rho_{cp}^2\gamma^d\ell^{2d}}{2} \sum_{i,j\leq \bfn}[A_{i,j}+B_{i,j}]
.\label{Delta1}\end{equation}
By assumption \eqref{eq:assphi3}, $\lim_{\gamma\to 0}(\gamma\ell)^d\sum_{i}B_{i,j}=O(1/m)$ uniformly in $j$. Analogously, by assumptions \eqref{eq:assphi2} and \eqref{eq:assphi3}, $(\gamma\ell)^d\sum_{i}A_{i,j}=O(\gamma\ell)$.
Putting things together, and recalling that $\bfn \ell^d=L^d$, we find
\begin{equation}\lim_{\gamma\to 0}\lim_{L\to\infty}|\Delta(\bfN)| L^{-d}=O(1/m).\label{Delta2}\end{equation}
Since the second term in the the right-hand side of \eqref{eq:Z1}, divided by $L^d$, vanishes as $L\to\infty$ then $\gamma\to 0$, using Stirling's formula, \eqref{eq:Zinf1} implies
\begin{equation}
\label{eq:Energy1}
\mathcal{F}_\beta(\rho)\leq 
\lim_{m\to\infty}\lim_{\ell\to\infty }\lim_{\gamma\to 0}\lim_{L\to\infty} \inf_\bfN A_\beta(\bfN),\end{equation}
where the infimum is carried out over all families $\bfN =(N_i^p)_{i\leq \bfn, p\leq \bfm}$ summing to $\rho L^d$ such that $N_i\leq \rho_{cp}\ell^d$, and 
\begin{equation}A_\beta(\bfN)=\frac{1}{\beta L^d}\sum_{i,p}N_i^p\log\frac{N_i^p}{N_i|\S_1|}\\
- \frac{1}{ \beta L^d}\sum_{i}  \log Z_{hc}(N_i,k) +\frac{\gamma^d}{2 L^d} \sum_{\substack{i,j\leq \bfn\\
p,q\leq \bfm}}N_i^pN_j^q\underline{\varphi}_{i,j}^{p,q}.
\end{equation}

\medskip

Fix $\tau>0$, and recall from Theorem \ref{thm:FreeEnergy} the definition of the set $ \mathscr{R}_\tau$. We are now ready to investigate the limit of the quantity 
above. 
Fix $\rho'<\rho_{cp}$ and $f\in  \mathscr{R}_\tau$ such that $\rho \bar f(x)\le \rho'$
(recall the notation $\bar f(x)=\int_{\S^2}f(x,\Omega)d\Omega$), and define $\widetilde{\bfM}=(\widetilde M_i^p)_{i\leq \bfn, p\leq \bfm}$ as
\begin{equation}
\label{eq:defMf}
\widetilde{M}_i^p=\left \lfloor\rho \int_{\Lambda^i_L\times \S_p} f(\gamma x,\Omega)dx d\Omega\right\rfloor.
\end{equation}
Note that we will take the limit $m\to\infty$ after $\ell\to\infty$, so that each of the $\widetilde{M}_i^p$ will go to $\infty$ in any cell where $f>0$. We then define the family $M_i^p$ by adding $1$ to the smallest $\widetilde{M}_i^p$'s in order to enforce $\sum_{i,p}M_i^p=\lfloor\rho L^d\rfloor$. We now investigate the limit of the right-hand side of \eqref{eq:Energy1} for $\bfN=\bfM$. Following the same steps as in \cite{GP1}, one can check that in the quadruple limit of \eqref{eq:Energy1}, since $\rho\bar f\leq\rho'<\rho_{cp}$,
\begin{eqnarray}
&&\bigg|\frac{1}{\beta L^d}\sum_{i,p}M_i^p\log\frac{M_i^p}{M_i|\S_1|}\\
&&\qquad - \frac{\rho }{\beta(L')^d}\int_{\Lambda_{L'}}\cro{\int_{\S^2}f(x,\Omega)\log(f(x,\Omega))d\Omega -\bar f(x)\log \bar f(x)}dx\bigg|\to 0.\nonumber
\end{eqnarray}
where we shortened $L'=\gamma L$. Moreover, by the periodicity of $f$, 
\begin{eqnarray} &\phantom{=}& \frac1{(L')^d}\int_{\Lambda_{L'}}\cro{\int_{\S^2}f(x,\Omega)\log(f(x,\Omega))d\Omega -\bar f(x)\log \bar f(x)}dx=\nonumber\\
&=&\frac1{\tau^d}\int_{\Lambda_\tau}\cro{\int_{\S^2}f(x,\Omega)\log(f(x,\Omega))d\Omega -\bar f(x)\log \bar f(x)}dx+O(1/L').\nonumber\end{eqnarray}
Similarly,
\begin{equation*}
\bigg|- \frac{1}{ \bfn}\sum_{i} \frac{1}{\beta \ell^d} \log Z_{hc}(M_i,k)- \frac{1}{(L')^d}\int_{\Lambda_{L'}}\mathcal{F}_{hc}(\rho \bar f(x))dx\bigg|\to 0,
\end{equation*}
and, by periodicity again, $\frac{1}{(L')^d}\int_{\Lambda_{L'}}\mathcal{F}_{hc}(\rho \bar f(x))dx=
\frac{1}{\tau^d}\int_{\Lambda_{\tau}}\mathcal{F}_{hc}(\rho \bar f(x))dx$. 
Finally 
\begin{multline*}
\bigg|\frac{\gamma^d}{2 L^d} \sum_{\substack{i,j\leq \bfn\\
p,q\leq \bfm}}M_i^pM_j^q\underline{\varphi}_{i,j}^{p,q}\\
-\frac{ \rho^2}{2(L')^d}\int_{\Lambda_{L'}\times\Lambda_{L'}}dx\, dx'\int_{\S^2\times \S^2}d\Omega\,d\Omega'\, f(x,\Omega)f(x',\Omega')\varphi(x-x', \Omega\cdot\Omega') \bigg|\to 0.
\end{multline*}
In the last line, the integral over $\Lambda_{L'}\times\Lambda_{L'}$ can be freely replaced by $\Lambda_{L'}\times \mathbb R^d$, because 
$\rho\bar f$ is bounded by $\rho'$ and $\varphi $ is integrable. Finally, by periodicity, we can rewrite 
\begin{multline*}\frac{ \rho^2}{2(L')^d}\int_{\Lambda_{L'}\times\mathbb R^d}dx\, dx'\int_{\S^2\times \S^2}d\Omega\,d\Omega'\, f(x,\Omega)f(x',\Omega')\varphi(x-x', \Omega\cdot\Omega')\\
=\frac{ \rho^2}{2\tau^d}\int_{\Lambda_{\tau}\times\mathbb R^d}dx\, dx'\int_{\S^2\times \S^2}d\Omega\,d\Omega'\, f(x,\Omega)f(x',\Omega')\varphi(x-x', \Omega\cdot\Omega').
\end{multline*} 

\medskip

To conclude, let us explicitly write $\bfM=\bfM(\rho,f)$ to indicate the dependency of $\bfM$ on $\rho$ and on $f \in \mathscr{R}_\tau$. Define $\Gamma_{\bfn,\bfm}^\rho(\rho')$ (resp. $ \mathscr{R}_\tau(\rho')$) the set of $\bfN$'s that sum up to $\rho L^d$ and such that each of the $N_i$'s is bounded by $\rho'\ell^d$ (resp. functions $f\in \mathscr{R}_\tau$ such that $\rho\bar f$ is bounded by $\rho'<\rho_{cp}$). We rewrite  \eqref{eq:Energy1} as
\begin{eqnarray}
\mathcal{F}_\beta(\rho)&
\leq& \lim_{\rho'\to\rho_{cp}} \lim_{m\to\infty}\lim_{\ell\to\infty }\lim_{\gamma\to 0}\lim_{L\to\infty} \inf_{\bfN\in \Gamma^\rho_{\bfn,\bfm}(\rho')} A_\beta (\bfN)\label{eq:upperfin}\\
&\leq& \lim_{\rho'\to \rho_{cp}}\cro{\lim_{m\to\infty}\lim_{\ell\to\infty }\lim_{\gamma\to 0}\lim_{L\to\infty} \inf_{\substack{\tau>0\\ f\in \mathscr{R}_\tau(\rho')}} A_\beta(\bfM(\rho,f))},\nonumber
\end{eqnarray}
which, thanks to the three estimates above, proves the upper bound after straightforward computations, since for $\rho'<\rho_{cp}$ fixed, 
we can exchange the limits in brackets with the inf.

\subsection{Lower bound}

For the lower bound on the free energy, which corresponds to an upper bound on the partition function, we first write
\begin{equation}\label{yeta}Z_{\beta}(N,L,\gamma )\le\sum_{\bfN}\left[\frac{1}{\prod_{i,p}N_i^p!}\int_{( \Lambda^1_L\times \S_1)^{N_1^1}}\dots\int_{( \Lambda^\bfn_L\times \S_\bfm)^{N_{\bfn}^\bfm}}d\bar x d\bar\Omega e^{-\beta W_{\min}(\bfN)}\right],\end{equation}
where  the sum is taken over all families $\bfN$ summing to $N$; moreover, $W_{\min}(\bfN)$ is a lower bound on $V_\gamma$ for configurations with $N_i^p$ particles in $\Lambda_i\times \S_p$, of the same form as \eqref{eq:Z1}, with the only difference that $\Delta(\bfN)$ is replaced by a different the remainder $\Delta'(\bfN)$, bounded in the same way as \eqref{Delta1}-\eqref{Delta2}. From \eqref{yeta}, we get
\begin{equation}
\label{eq:Zinf2}
Z_{\beta}(N,L,\gamma )\leq 2^{\bfm}(\rho_{cp}\ell^d)^{\bfn}\sup_{\bfN}\left[|\S_1|^{N}\frac{\prod_{i}N_i!}{\prod_{i,p}N_i^p!}\prod_{i}Z_{hc}(N_i,\ell) e^{-\beta W_{\min}(\bfN)}\right],
\end{equation}
where the factor $2^{\bfm}(\rho_{cp}\ell^d)^{\bfn}$ in the right-hand side is a crude upper bound on the number of families $(N_i^p)$ such that $\sum_p N_i^p=N_i$ 
and each $N_i$ is less than $\rho_{cp}\ell^d$; to obtain this bound, we simply ignored the constraint that  $\sum_i N_i=N$. Note that, in the limit
\eqref{eq:quadruplelimit}, the contribution of this term to the free energy vanishes. Therefore, by repeating the same considerations as in the previous subsection, we
find that the free energy $\mathcal F_\beta(\rho)$ is bounded from below by the same expression as the right side of \eqref{eq:upperfin}, and this concludes the proof
of the theorem. 

\section{Phase transition in liquid crystals: proof of Theorem \ref{thm:PhaseTransition}}
\label{sec:PhaseTrans}

We recall that the functional of interest is the $r_0\to 0^+$ limit of \eqref{eq:freeenergyfun}, namely 
\begin{eqnarray}
 \mathcal{F}^0_{\beta,\tau}(\rho,f)&=&\frac1{\tau^d}\bigg\{\frac{\rho}{\beta}\int_{\Lambda_\tau\times \S^2} f(x,\Omega)\log  f(x,\Omega)  dx\, d\Omega \label{eq:freeenergyfun0}\\
&+&\frac{\rho^2}{2} \int_{\Lambda_\tau\times \R^d}dx\,dy\int_{\S^2\times \S^2}d\Omega\, d\Omega' f(x,\Omega)f(y, \Omega')\varphi(x-y,\Omega\cdot\Omega') \bigg\},\nonumber
\end{eqnarray}

\paragraph{{\bf Proof of item 1.}} It is straightforward to check that $f_0\equiv 1/4\pi$ 
is a critical point of $\mathcal{F}^0_{\beta,\tau}(\rho,f)$, for all $\beta, \rho,\tau$. In order to show that $f_0$ is the global minimizer for $\rho$ small enough, 
fix $\tau\ge 0$ and write $f\in \mathscr{R}_\tau$ as $f=f_0(1+4\pi h)$, with $\int_{\Lambda_\tau\times\S^2} h=0$ and $4\pi h\ge -1$. 
We have:
\begin{equation} \mathcal{F}^0_{\beta,\tau}(\rho,f)-\mathcal{F}^0_{\beta,\tau}(\rho,f_0)=
\frac{\rho}{\beta}\langle (1+4\pi h)\log (1+4\pi h) \rangle +\mathcal E_\tau(\rho, h), \end{equation}
where 
$\langle F\rangle=\frac1{4\pi\tau^d}\int_{\Lambda_\tau\times\S^2}F(x,\Omega)dx\,d\Omega$ 
and 
\begin{multline*}
\mathcal{E}_\tau(\rho,h)=\frac{\rho^2}{2\tau^d} \int_{\Lambda_\tau\times \R^d}dx\,dy\int_{\S^2\times \S^2}d\Omega\,d\Omega' \ h(x,\Omega) h(y, \Omega')
\varphi(x-y,\Omega\cdot\Omega').
\end{multline*}
Thanks to the periodicity of $h(\cdot,\Omega)$, we can rewrite 
\begin{multline*}
\mathcal{E}_\tau(\rho,h)=\frac{\rho^2}{2\tau^{d}} \int_{\Lambda_\tau\times \Lambda_\tau}dx\,dy\int_{\S^2\times \S^2}d\Omega\,d\Omega' \ h(x,\Omega) h(y, \Omega')
\varphi_\tau(x-y,\Omega\cdot\Omega'),
\end{multline*}
where $\varphi_\tau(x,u)=\sum_{n\in\mathbb Z^d}\varphi(x+n\tau,u)$. Now, thanks to the integrability condition \eqref{eq:supsumbound}, 
$|\varphi_\tau(x,u)|\le K \tau^{-d}$ for some constant $K$, uniformly in $\tau, u$. Therefore, 
\begin{equation} \mathcal{F}_{\beta,\tau}^0(\rho,f)-\mathcal{F}_{\beta,\tau}^0(\rho,f_0)\ge 
\frac{\rho}{\beta}\langle (1+4\pi h)\log (1+4\pi h) \rangle -8\pi^2\rho^2 K\langle  |h|\rangle^2.\label{case1.1} \end{equation}
Let $h_+$ and $h_-$ be the positive and negative parts of $h$, respectively. We let $H=4\pi\langle h_+\rangle=4\pi\langle h_-\rangle\le 1$. Using the convexity 
of $(1+x)\log(1+x)$ and of $(1-x)\log(1-x)$, we find
$$\langle (1+4\pi h)\log (1+4\pi h) \rangle \ge (1+H)\log(1+H)+(1-H)\log(1-H),$$
which is bounded from below by $H^2$, for all $0\le H\le 1$. Using also the fact that $4\pi\langle |h|\rangle=2H$, from \eqref{case1.1} we find 
\begin{equation} \mathcal{F}_{\beta,\tau}^0(\rho,f)-\mathcal{F}_{\beta,\tau}^0(\rho,f_0)\ge 
\frac{\rho}{\beta}H^2  - 2\rho^2 K H^2, \end{equation}
which proves that $f_0$ is the unique global minimizer, if $\rho<(2\beta K)^{-1}$. This proves item 1 of Theorem \ref{thm:PhaseTransition}, with $\rho_c$ the sup 
of the values of $\rho$ for which $f_0$ is the unique global minimizer. 

\medskip

It is easy to see that at any density beyond $\rho_c$ there exists a non-uniform $f$ with lower free energy than $f_0$. To see this, for any $\epsilon>0$, choose 
$\rho_c<\rho\le \rho_c+\epsilon$, in correspondence of which there is $\tau\ge 0$ and $f\neq f_0$ in $\mathscr{R}_\tau$ such that 
\begin{equation}\label{5.1}\mathcal{F}_{\beta,\tau}^0(\rho,f)\le \mathcal{F}_{\beta,\tau}^0(\rho,f_0).\end{equation} 
This is possible by the very definition of $\rho_c$. More explicitly, \eqref{5.1} means 
\begin{eqnarray}\label{5.2}
&&\int_{\Lambda_\tau\times \S^2}\Big( f(x,\Omega)\log  f(x,\Omega) -f_0\log f_0\Big) dx\, d\Omega \le \\
&&\qquad \le 
\frac{\beta\rho}{2} \int_{\Lambda_\tau\times \R^d}dx\,dy\int_{\S^2\times \S^2}d\Omega\, d\Omega' \varphi(x-y,\Omega\cdot\Omega')\Big(f_0^2-
f(x,\Omega)f(y, \Omega')\Big).\nonumber
\end{eqnarray} 
The left side is positive, because $f\log f$ is strictly convex and $f$ is non uniform. As a consequence, the right side is positive, too; so, if we take any $\rho'>\rho$,
\begin{eqnarray}\label{5.3}
&&\int_{\Lambda_\tau\times \S^2}\Big( f(x,\Omega)\log  f(x,\Omega) -f_0\log f_0\Big) dx\, d\Omega \\
&&\qquad <
\frac{\beta\rho'}{2} \int_{\Lambda_\tau\times \R^d}dx\,dy\int_{\S^2\times \S^2}d\Omega\, d\Omega' \varphi(x-y,\Omega\cdot\Omega')\Big(f_0^2-
f(x,\Omega)f(y, \Omega')\Big),\nonumber
\end{eqnarray} 
that is, $\mathcal{F}_{\beta,\tau}^0(\rho',f)< \mathcal{F}_{\beta,\tau}^0(\rho',f_0)$, as announced. 

\medskip

\paragraph{{\bf Proof of items 2 and 3.}} In light of what we already proved above, in order to complete the proof of item 2 of Theorem \ref{thm:PhaseTransition}, 
we are left with proving that the value of the density at which $f_0$ loses linear stability is strictly larger than $\rho_c$. Fix $\tau\ge 0$. 
Consider a small perturbation $f=f_0+\varepsilon h$ , where $h$ is $\tau$-periodic such that $\int_{\Lambda_\tau\times \S^2} h=0$. If we expand the free energy 
up to order $\varepsilon^3$ included we find: 
\begin{eqnarray}
&&\mathcal F_{\beta,\tau}^0(\rho,f_0+\varepsilon h)-\mathcal F_{\beta,\tau}^0(\rho,f_0)=\label{5.8}\\
&&\qquad =
\varepsilon^2\Big[\frac{8\pi^2 \rho}{\beta}\langle h^2\rangle +\mathcal{E}_\tau(\rho,h)\Big]-
\varepsilon^3\frac{32\pi^3\rho}{\beta}\langle h^3\rangle+O(\varepsilon^4).\nonumber
\end{eqnarray}
As the next step, we diagonalize $\mathcal E_\tau(\rho,h)$. Passing to Fourier space with respect to the $x$ variable, we get 
\begin{equation}
\label{eq:FirstTransform}
\mathcal{E}_\tau(\rho,h)=\frac{\rho^2}{2\tau^{2d}}\sum_{k\in \Z^d}\int_{ \S^2\times \S^2} \hat h_k(\Omega) \hat h_{-k}(\Omega') \hat{\varphi}_k(\Omega\cdot\Omega')d\Omega d\Omega',
\end{equation}
where for $ k\in \Z^d$
\[\hat h_k(\Omega)=\int_{\Lambda_\tau}h(x, \Omega)e^{\frac{2i\pi }{\tau}k \cdot x}dx \quad \mbox{ and }\quad \hat \varphi_k(u)=\int_{\R^d }\varphi(y, u)e^{-\frac{2i\pi }{\tau}k \cdot y} dy.\]
Next, for any $k\in \Z^d$, define the operator $\mathcal{G}_k$ acting on a function $g$ on $\S^2$ as 
\[(\mathcal{G}_k g)(\Omega)=\int_{\mathbb{S}^2} \hat \varphi_k\big(\Omega\cdot \Omega') g(\Omega') d\Omega'.\]
For later reference, we denote by $G_k(\Omega,\Omega')=\hat \varphi_k(\Omega\cdot\Omega')$ the kernel of $\mathcal G_k$. 
Moreover, let $L_z, L_\pm$ be the usual angular momentum operators
\[L_z=\frac{1}{i}\frac{\partial}{\partial\phi}\quad \mbox{ and } \quad L_\pm=e^{\pm i\phi}\pa{\pm\frac{\partial}{\partial \theta}+i\cot(\theta)\frac{\partial}{\partial\phi}},\]
where $(\theta,\phi)\in [0,\pi]\times[0,2\pi]$ are the spherical coordinates of an element $\Omega\in \mathbb{S}^2$.
Straightforward computations, for $\Omega=( \theta, \phi)$ and $\Omega'=( \theta', \phi')$, yield
\[\Omega\cdot\Omega'=\cos\theta\cos\theta'+\sin\theta\sin\theta'\cos(\phi-\phi')\]
\begin{align*}
[L_\pm G_k(\cdot,\Omega')](\Omega)&=\partial_u \hat\varphi_k(\Omega\cdot\Omega')\cro{\mp \sin\theta \cos\theta' e^{\pm i\phi}\pm \cos \theta \sin\theta' e^{\pm i\phi'}}\\
&=-[L_\pm G_k (\Omega,\cdot)](\Omega').
\end{align*}
\[[L_z G_k(\cdot,\Omega')](\Omega)=\partial_u\hat \varphi_k(\Omega\cdot\Omega') \sin \theta \sin \theta' i \sin(\phi-\phi')=-[L_z G_k (\Omega,\cdot)](\Omega'),\]
which, by integration by parts, proves that $\mathcal{G}_k$ commutes both with $L_z$ and with $L_\pm$.

\medskip 

As defined in more details in Appendix \ref{app:Legendre}, we consider the spherical harmonics defined for $\ell \in \N_0$, $|m|\leq \ell$
\begin{equation*}
Y^m_{\ell}(\theta, \phi)=C_{\ell,m }P_{\ell}^m(\cos \theta)e^{im\phi} ,\end{equation*}
where the $C_{\ell,m }$ are normalizing constants making it an orthonormal family, and the $P_{\ell}^m$'s are the associated Legendre polynomials. They satisfy the classical relations 
\[L_\pm Y_{\ell}^m=\sqrt{(\ell\mp m )(\ell\pm m+1)}Y_{\ell}^{m\pm1}\] 
and
\[L^2 Y_{\ell}^m:=  [L_+L_-+L_z^2] Y_{\ell}^m=\ell(\ell+1)Y_{\ell}^m.\]
In particular, since $\mathcal{G}_k$ commutes with $L_z$ and $L_\pm$, it also commutes with the angular momentum $L^2$, therefore its eigenvectors are the $Y_\ell^m$'s. Furthermore, since $\mathcal{G}_k$ commutes with $L_\pm$, the corresponding eigenvalues $\lambda_{\ell,m}(k)=\lambda_\ell(k)$ do not depend on $m$. Writing the identity $\lambda_\ell(k) Y^0_\ell(\theta,\phi)=(\mathcal G_k Y^0_\ell)(\theta,\phi)$ at $\theta=0$, we get 
\[\lambda_\ell(k)=2\pi \int_{-1}^1P_\ell(u)\hat\varphi_k(u)du=2\pi  \hat\Phi_\ell\pa{\frac{2k\pi}{\tau}}.\] 
where $\hat \Phi_\ell(\xi)$ was defined in \eqref{eq:defPhi}. If we now expand each of the $\hat h_k$ appearing in \eqref{eq:FirstTransform} in spherical 
harmonics, we can rewrite
\begin{eqnarray}
\label{eq:SecondTransform}
\mathcal{E}_\tau(\rho,h)&=&\frac{\pi  \rho^2 }{\tau^{2d}}\sum_{\ell\in \N_0} \sum_{ m=-\ell}^\ell \sum_{k\in \Z^d} |\hat h_{k,\ell,m}|^2   \hat\Phi_\ell\pa{\frac{2k\pi}{\tau}}\\
&\geq& \frac{\rho^2 \hat \Phi_{\ell^\star}(0)}{2\tau^{d}} \int_{\Lambda_\tau\times \S^2}h^2(x,\Omega) dx d\Omega=
2\pi \rho^2\hat \Phi_{\ell^\star}(0)\langle h^2\rangle.
\end{eqnarray}
To establish the lower bound, we used  Assumption \ref{eq:cond_phi} and both Fourier and Spherical harmonics versions of Parseval's identity. 

Plugging this back into \eqref{5.8}, we see that the square brackets in the right side is bounded from below as
\begin{equation}\frac{8\pi^2 \rho}{\beta}\langle h^2\rangle +\mathcal{E}_\tau(\rho,h)\ge \frac{4\pi^2\rho}{\beta}
\big(2-(\beta\rho/2\pi)|\hat \Phi_{\ell^\star}(0)|\big)\langle h^2\rangle,\label{inequalityh}\end{equation}
which proves the linear stability of $f_0$ for any $\rho<\rho^\star:=\frac{4\pi }{\beta |\hat \Phi_{\ell^\star}(0)|}$. 
Let us identify $\Omega\in \S^2$ with its polar coordinates $(\theta, \phi)$, and let $h^\star$ be the function $h^\star(x,\theta,\phi):=
 P_{\ell^\star}(\cos\theta)$.
Note that, choosing $h=h^\star$, Eq. \eqref{inequalityh} is valid with the equality sign. This implies the linear stability of $f_0$ for any $\rho>\rho^\star$, 
thus proving item 3 of Theorem \ref{thm:PhaseTransition}. 

\medskip

We are left with proving that $\rho_c<\rho^\star$. For this purpose, compute \eqref{5.8} at $\tau=0$, $\rho=\rho^\star$ and $h=h^\star$: 
\begin{equation}\label{5.13}
\mathcal{F}_\beta^0(\rho^\star ,f_0+\varepsilon h^\star)-\mathcal{F}_\beta^0(\rho^\star ,f_0)=-\varepsilon^3\frac{32\pi^3\rho^\star}{\beta}\langle (h^\star)^3\rangle
+O(\varepsilon^4).\end{equation}
The $\varepsilon^3$ term in the right side equals 
$$-\varepsilon^3\frac{16\pi^3\rho^\star}{\beta} \int_{-1}^1 P_{\ell^\star}^3(u) du.$$
Recalling  that $\ell^\star$ is even, see Remark \ref{rem:elleven}, this integral can be computed explicitly, recognizing that it is a special case of Gaunt's formula, see e.g. \cite[(7.125), p.771]{GR}; we thus get
\[\int_{-1}^1 P_{\ell^\star}^3(u)du
=2\frac{(2s)!^3}{(6s+1)!}\frac{(3s)!^2}{s!^6}>0\]
where we used the shorthand notation $s$ for the positive integer $s=\ell^\star/2$. 
In particular, for $\varepsilon$ small enough, the r.h.s. of \eqref{5.13} is negative, which proves that at $\rho=\rho^\star$, the uniform profile is not a global minimizer of $\mathcal{F}_\beta^0(\rho,f)$. This concludes the proof of Theorem \ref{thm:PhaseTransition}.

 \begin{remark}[Case of the angular dimension $2$, $\Omega\in \S^1$]
\label{rem:d2}
Note that the same analysis can be applied to study the free energy functional for angles in $\S^1$. In this case, the density $\rho^\star$ at which the uniform profile loses linear stability can be expressed as a function of the minimal Fourier coefficient, in orientation \emph{and} space, of $\varphi$.
 In this case however, $\int \cos^3(u)du$ vanishes, and the next contribution $O(\varepsilon^4)$ in \eqref{5.13} is positive. One therefore expects that the transition is instead continuous, i.e. that for any $ \rho<\rho^\star$, the uniform profile is the global minimizer. As mentioned in the introduction, this has in fact been proved for 
 some special choices of the interaction potential, including Maier Saupe's, see \cite{FS2}, but to the best of our knowledge a proof for (more) general potentials is missing.   
 \end{remark}
 
\begin{remark}[Magnetic interaction vc. Liquid crystals]
\label{rem:ferromag}
The choice of a magnetic rather than liquid-crystalline interaction formally corresponds to a choice of an odd function $\varphi(x,\cdot)$. In this case, $\lambda_\ell(k)$ vanishes for $\ell$ even instead. However, for any $\ell$ odd, the integral $\int_{\mathbb{S}^2}P_{\ell}^3$ vanishes, so that the proof above for a discontinuous phase transition no longer holds. In fact, the converse holds, and at the critical point $\rho^\star$, for an odd interaction potential $\varphi(x,\cdot)$, for $\varepsilon$ small enough,
\[\mathcal{F}_\beta(\rho^\star ,f_0+\varepsilon h^\star)\geq \mathcal{F}_\beta(\rho^\star ,f_0).\]
This is in line with the known fact that the mean field Heisenberg model undergoes a second order phase transition \cite{KN}; an analogous fact is expected 
for more general magnetic interactions. 
\end{remark}

\appendix

\section{Legendre polynomials and spherical harmonics}
\label{app:Legendre}

We first recall some basic properties of the \emph{Legendre polynomials} $P_\ell$. For more on the topic, we refer the reader to e.g. \cite{AW}. This is the unique family of polynomials $P_\ell:[-1,1]\to \R$ satisfying 
\begin{itemize}
\item [--] for any $\ell\in \N_0$, $P_\ell$ is a degree $\ell$ polynomial and $P_\ell(1)=1$;
\item [--] the family $(P_\ell)_{\ell\in \N_0}$ is orthonormal in $L^2$, i.e. $\int_{-1}^1P_{\ell}(u)P_k(u)du =\frac{2}{2\ell+1}{\bf 1}_{\{\ell=k\}}$ for any $k, \ell \in \N_0$.
\end{itemize}
In particular, since $P_0\equiv 1$, for any $\ell \neq 0$, $\int_{-1}^1P_{\ell}(u)du =0$. The Legendre polynomials have same parity as $\ell$. The Legendre polynomials satisfy the Bonnet's recursion formula 
\begin{equation}
\label{eq:Bonnet}
(n+2)P_{n+2}=(2n+3)uP_{n+1}-(n+1)P_n.
\end{equation}

\bigskip

The \emph{associated Legendre polynomials $P_\ell^m$ } can then be defined for $m=0,\dots, \ell$ as
\begin{equation}\label{assLegpol}P_\ell^m(u)=(-1)^m(1-u^2)^{m/2}\frac{d^m}{du^m}[P_\ell(u)],\end{equation}
 and for $m=-1,\dots, -\ell$, 
\[P_\ell^{-m}(u)=(-1)^m\frac{(\ell-m)!}{(\ell+m)!}P_\ell^m(u).\]
Those polynomials satisfy the orthogonality relations 
\[\int_{-1}^1P^m_{\ell}(u)P^m_k(u)du =\frac{2(\ell+m)!}{(2\ell+1)(\ell-m)!}{\bf 1}_{\{\ell=k\}}\]
and
\[\int_{-1}^1\frac{P^m_{\ell}(u)P^n_\ell(u)}{1-u^2}du =
\frac{(\ell+m)!}{m(\ell-m)!}{\bf 1}_{\{m=n>0\}}+\infty \times {\bf 1}_{\{m=n=0\}}.\]
The associated Legendre polynomials have the same  parity as $\ell+m$, 
\[P_{\ell}^m(-u)=(-1)^{\ell+m}P_\ell^m(u).\]

\bigskip

The \emph{spherical harmonics} $Y_\ell^m$ can finally be defined as the functions  $Y_\ell^m:\S^2\to \R$
\[Y_\ell^m(\theta, \phi)=C_{\ell,m}P_{\ell}^m(\cos \theta)e^{im\phi},\]
where $C_{\ell,m}=\sqrt{\frac{(2\ell+1)(\ell-m)!}{4\pi(\ell+m)!}}$ are normalizing constants making the family orthonormal, 
\[\int_{\theta=0}^\pi\int_{\phi=0}^{2\pi}Y_\ell^m(\theta, \phi)Y_k^{n,*}(\theta, \phi)\sin\theta d\theta d\phi={\bf 1}_{\{k=\ell, \;n=m\}}.\]

\bigskip

{\bf Acknowledgements}: We would like to thank Michele Correggi and Ian Jauslin for very fruitful discussions.
This work has been supported by the European Research Council (ERC) under the European Union's Horizon 2020 research and innovation programme (ERC CoG UniCoSM, grant agreement n.724939). A.G. would also like to acknowledge financial support from MIUR, PRIN 2017 project MaQuMA cod. 2017ASFLJR.

\bigskip

 {\bf Data availability statement}: Data sharing is not applicable to this article as no new data were created or analyzed in this study.

\end{document}